\documentclass[preprint,onecolumn]{elsarticle}
\usepackage{inputenc}
\usepackage{braket}
\usepackage{hyperref}
\usepackage[super]{cite} 
\usepackage{amsmath}
\usepackage{xcolor}
\usepackage{comment}

\begin{document}


\title{3D polymer simulations of genome organization and transcription across different  chromosomes and cell types }

\author{Massimiliano Semeraro}
\address{Dipartimento di Fisica, Universit\`a degli Studi di Bari, and INFN, Sezione di Bari, via Amendola 173, 70126 Bari, Italy\\
massimiliano.semeraro@uniba.it}

\author{Giuseppe Negro}
\address{SUPA, School of Physics and Astronomy, University of Edinburgh, Peter Guthrie Tait Road, Edinburgh EH9 3FD, UK \\  Dipartimento di Fisica, Universit\`a degli Studi di Bari, and INFN, Sezione di Bari, via Amendola 173, 70126 Bari, Italy\\
giuseppe.negro@ba.infn.it}

\author{Antonio Suma}
\address{Dipartimento di Fisica, Universit\`a degli Studi di Bari, and INFN, Sezione di Bari, via Amendola 173, 70126 Bari, Italy\\
antonio.suma@uniba.it}

\author{Giuseppe Gonnella}
\address{Dipartimento di Fisica, Universit\`a degli Studi di Bari, and INFN, Sezione di Bari, via Amendola 173, 70126 Bari, Italy\\
giuseppe.gonnella@ba.infn.it}

\author{Davide Marenduzzo}
\address{SUPA, School of Physics and Astronomy, University of Edinburgh, Peter Guthrie Tait Road, Edinburgh EH9 3FD, UK\\ Davide.Marenduzzo@ed.ac.uk}

\begin{abstract}
We employ the {\it diffusing transcription factors model} for numerical simulation of chromatin topology conformations and transcriptional processes of human chromatin. Simulations of a short chromatin filament reveal different possible pathways to regulate transcription: it is shown that the transcriptional activity profile can be regulated and controlled by either acting on the chain structural properties, or on external factors, such as the number of transcription factors. Additionally, comparisons between GRO-seq experimental data and large scale numerical simulation of entire chromosomes from the human umbilical vein endothelial cell and the B-lmphocyte GM12878 show that the model provides reliable and statistically significant predictions for transcription across different cell-lines.
\end{abstract}

\maketitle



\section{Introduction}

In every human cell about $2$ meters of DNA filaments are packed inside a nucleus which is only $\sim10$ micrometers in diameter \cite{2meters}. In order to characterise the DNA chromatin degree of spatial organization in the nucleus and its connections with the vital transcription, replication and repair cellular functions, several advanced experimental techniques, such as, just to mention a few, ``Hi-C'' and super resolution microscopy\cite{3C_review,Marko2021,Lafontaine2021,Birk2019}, have been developed. Such novel experimental tools proved fundamental in revealing complex chromatin topological features as three-dimensional (3D) conformation at different length scales. At large scales, interactions between active regions were discovered to be the driver of the genome segregation into compartments\cite{compartments}; at shorter scales, chromosomes were shown to be partitioned into contiguous regions termed domains, or topologically associating domains (TADs) \cite{tads}, characterised by enriched self-interactions and chromatin loops\cite{loops}.

In order to better frame these {\it in vitro} experimental data, in recent years much effort was put into the theoretical development of a number of polymer and statistical physics models for DNA and chromatin\cite{NICODEMI201490,bukle2018,marenduzzoreview,Bianco2017,nicodemi2022,Esposito2021,henrich2018,henrich2022,rao2022,rao2013,foglino2019}, able to numerically reproduce {\it in silico} the stochastic organization of genomic loci or even entire chromosomes \cite{NICODEMI201490, bukle2018, marenduzzoreview, Bianco2017}. Two are the commonly adopted approaches: inverse, or fitting-based, models and forward, or mechanistic, models. The inverse modelling approach is top-down: one starts from experimental data and relies on sophisticated fitting procedures to infer most possible realistic polymer models \cite{Zhan2017, Bianco2017}. On the opposite, the forward modelling approach is bottom-up \cite{Bianco2017,Habbad2017}: one starts from few reasonable biological and biophysical assumptions to produce computer simulations outputs directly comparable to as many types of experimental data as possible. 

In combination with experimental evidence, the use of forward models resulted instrumental in providing mechanistic models for genome organization. Three are the main polymer models offered by the literature: the block copolymer model \cite{Jost2014,MichielettoPhysRevX041047}, the loop extrusion model \cite{extrusion1,extrusion2,extrusion3,extrusion4} and the diffusing transcription factors model \cite{brackley2013, brackley2020b,brackley2021, bukle2018, brackley2016}. In the block copolymer model, each polymer bead coarse grains a chromosome region and is assigned a specific species. Beads of the same type   interact attractively so as to model direct chromatin-chromatin interactions. The loop extrusion model instead assumes the existence of active factors, or extruders, able to bind the chromatin filament at a single point and generate a growing loop by tracking along the contour of the fibre.

In this article, we focus on the the diffusing transcription factors model, in which chromosomes are modelled as bead-and-spring polymers, each bead resolving $\sim 10^{3}$ consecutive chromosome base pairs. The polymer beads modelling the promoter or enhancer regions interact attractively with transcription factors, i.e. free beads modelling proteins generically associated with the transcription process of DNA into messenger RNA. One of the main features of the diffusing transcription factors model is the so called \textit{bridging-induced attraction} feedback mechanism driving the formation and stability of clusters \cite{brackley2016, brackley2021,brackley2017}. The latter is a fundamental ingredient allowing the model to consistently reproduce Hi-C maps of {\it Drosophila}~\cite{michieletto2016} and human chromosomes \cite{brakley2013, brackley2016}, and to highlight the emergence of a small-world gene regulatory network driven by 3D structure\cite{brackley2021}. 
Although both experimental Hi-C maps and transcriptional GRO-seq data could be compared to numerical data produced by the diffusing transcription factors model, extensive comparisons have been performed in the former case but not in the latter. 

The aim of the present paper is to extensively test the predictive power of the diffusing transcription factors model. First, we provide a physical characterisation of the model by simulating a short fictitious chromatin filament to show how the structural properties of the polymer chain, i.e. the specific sequence of beads types, and external factors, such as the average number of active transcription factors, affect the transcriptional activity profile. Second, we perform extensive numerical simulations of three different chromosomes, HSA $14$, $18$ and $19$, on two different cell-lines, the human umbilical vein endothelial cell and the B-lymphocyte GM12878, to outline how the model performs compared to GRO-seq experimental data.

The structure of the paper is as follows. In \autoref{sec:model} we detail the model, provide technical details on simulations and present the mapping between reduced and physical units. \autoref{sec:results} is completely devoted to the results of our investigation. In particular, in \autoref{sec:toy_model} we consider the short chromatin filament, while in \autoref{sec:chrs} we report and comment the results concerning chromosome simulations. Finally in \autoref{sec:conclusions} we summarise the conclusions of our investigation.

\section{MODEL and METHODS}
\label{sec:model}

The aim of the present section is to introduce the {\it diffusing transcription factors model} and report simulations details.

\subsection{Diffusion transcription factors model}

Chromatin fibres are coarse grained into bead-and-spring polymer chains consisting of $M$ monomers. Each monomer represents $3000$ base-pairs (bp), and has a diameter of $\sigma\simeq30$ nm \cite{brackley2013}. The {\it transcription factors} (TFs) are modelled as $N$ beads of the same diameter which can bind differently to three monomer types of the DNA chain. Along the polymer there are in fact $n_{TU}$ high affinity {\it transcription units} (TUs), $n_w$ low affinity (weakly-sticky) beads, and $n_{nb}$ non-binding polymer beads, so that $M=n_{nb}+n_w+n_{TU}$. TFs can be in either an active binding or an inactive non-binding state, and are able to switch state at fixed time intervals with switching off and on rates $\alpha_{off}$ and $\alpha_{on}$, respectively, so that on average there are $N_a$ active state TFs and $N_p$ passive ones, with $N=N_a+N_p$. Active state TFs interact attractively strongly to TUs, weakly to weakly-binding beads and repulsively with non-binding beads, while passive state TFs interact repulsively with all polymer beads. Adopting a discrete index $i$ running from $1$ to $N+M$ for both TFs and polymer beads ,
$\textbf{r}_i=(r_{ix},r_{iy},r_{iz})$ will denote the position of the $i$-th generic bead in the three-dimensional space, while $r_{ij}=\sqrt{\sum_{\alpha=x,y,z}(r_{i\alpha}-r_{j\alpha})^2}$ the Euclidean distance between the $i$-th and the $j$-th beads.
 
Concerning the polymer beads, the generic $i$-th and $j$-th beads interact purely repulsively via the Weeks-Chandler-Anderson potential:
\begin{equation}
U^{ij}_{WCA}=\begin{cases}4k_BT\left[\left(\frac{\sigma}{r_{ij}}\right)^{12}   - \left(\frac{\sigma}{r_{ij}}\right)^{6} +\frac{1}{4}  \right]&\text{~if~}r_{ij}<2^{1/6}\sigma\\
0 & \text{otherwise}
\end{cases}~,
\end{equation}
where $k_B$ is the Boltzmann constant and $T$ is the environment temperature. In order to enforce the bead connectivity, any two consecutive chain beads interact through the harmonic potential
\begin{equation}
U^{ij}_{harmonic}=-k_h\left(r_{ij}-\bar{r}\right)^2~,
\end{equation}
 where $k_h = 100 k_B T/\sigma^2$ is the harmonic constant and $\bar{r}$ is the equilibrium spring distance set at $\bar{r}=1.1~\sigma$ \cite{brackley2021}. Finally, the stiffness of the chromatin fibre is taken into account by making each three consecutive beads along the polymer interact via the Kratky-Porod potential
 \begin{equation*}
U^{ij}_{KP}=\frac{k_B T l_P}{\sigma}\left[1-\frac{\vec{s_i}\cdot\vec{s_j}}{|\vec{s_i}||\vec{s_j}|}\right]=k_{kp}\left[1-\cos(\theta)\right]~,
 \end{equation*}
where $i$ and $j$ are neighbouring beads, $\vec{s_i}$ is the tangent vector connecting the $i$-th and the $i+1$-th beads, $\theta$ is the angle formed by such tangents and and $l_P$ is the persistence length of the chain.

Concerning polymer beads and TFs excluded volume interactions, any polymer bead, say the $a$-th, and TF, say the $b$-th, interact via the following truncated and shifted Lennard-Jones potential,
\begin{equation*}
U^{ab}_{LJ}=
\begin{cases}
4\epsilon_{ab}\left[\left(\frac{\sigma}{r_{ab}}\right)^{12}-\left(\frac{\sigma}{r_{ab}}\right)^6-\left(\frac{\sigma}{r_c}\right)^{12}+\left(\frac{\sigma}{r_c}\right)^6\right]&\text{~if~}r_{ab}<r_c\\
0 &\text{otherwise}
\end{cases}~,
\end{equation*}
where $r_c$ is the interaction cut-off set to $r_c=2^{1/6}\sigma$ for active and inactive TFs and non-binding polymer beads and for inactive TFs and weakly-sticky beads and TUs (purely repulsive interactions) or to $r_c=1.8\sigma$ for all active TFs and weakly-sticky beads and TUs (attractive interaction to model binding). The $U_{LJ}$ potential is shifted to zero at the cut-off. We set $\epsilon_{ab}=k_BT$ for purely repulsive interactions, $\epsilon_{ab}=3k_BT$ for attractive interactions between weakly-sticky beads and active TFs,  and $\epsilon_{ab}=8k_BT$ for strong attractive interactions between active TFs and TUs. A TU bead is considered transcriptionally active when it is bound to almost one TF, that is when the centres of the TU and an active TF are closer than a fixed distance threshold.

Finally the global time evolution of the system is described by the solution of the following system of $3(N+M)$ Langevin equations
\begin{equation}
    m_i\frac{d^2 r_{i\alpha}}{dt^2}=-\nabla U_i-\gamma_i\frac{dr_{i\alpha}}{dt}+\sqrt{2k_BT\gamma_i}~\eta_{i\alpha}(t)~,
\label{eq:system_of_eqs}
\end{equation}
where $i$ is the bead index running from $1$ to $N+M$, $\alpha=x,y,z$ is the dimension index, $m_i$ and $\gamma_i$ are the mass and friction coefficient associated to the $i$-th bead, $U_i$ is the total potential experienced by the $i$-th bead and $\eta_{i\alpha}(t)$ are a set of independent zero-mean delta-correlated stochastic white noises
\begin{equation*}
\braket{\eta_{i\alpha}(t)}=0~,\qquad\braket{\eta_{i\alpha}(t)\eta_{j\beta}(t')})=   \delta_{ij}\delta_{\alpha\beta}\delta(t-t')~,
\end{equation*}
where $\delta_{ij}$ and $\delta_{\alpha\beta}$ are two Kronecker deltas over the bead and three-dimensional indices, respectively, and $\delta(t-t')$ is a Dirac delta. For the sake of simplicity we set the mass $m_i\equiv m$ and friction coefficient $\gamma_i\equiv \gamma$, equal for all beads.

\subsection{Simulation details}

Coarse grained Brownian molecular dynamics (MD) simulations are performed using the LAMMPS (Large-scale Atomic/Molecular Massively Parallel Simulator) software package \cite{lammps}. We use  the Velocity-Verlet algorithm with the addition of the Langevin  thermostat modelling the stochastic thermal fluctuations and viscosity of an implicit solvent. For the sake of simplicity, hydrodynamic interactions are neglected and mass $m$, diameter $\sigma$, temperature $T$, and Boltzmann constant $k_B$ are all set to unity. The typical timescales are the Lennard-Jones time $\tau_{LJ}=\sigma\sqrt{m/\epsilon}$, with $\epsilon=k_BT$ the energy unit, and the brownian time $\tau_B=\sigma^2/D$, with $D=k_BT/\gamma$ the diffusion coefficient of a single bead. Setting $\gamma=1$, we have $\tau_B=\tau_{LJ}=1$.

The system of equations \autoref{eq:system_of_eqs} is integrated with timestep $\Delta t=0.01\tau_B$. The simulation domain is a cubic box with periodic boundary conditions of side $L$ chosen so that the system always results dilute. Protein switching is implemented by stochastically changing each TF state every $n_s$ timesteps as prescribed by the switching off and on rates $\alpha_{off}$ and $\alpha_{on}$. The actual box side length, TF number, TFs switching rates, $n_s$ and evolution time length will be specified when describing each case. 

Simulations begin with the chromatin fibre monomers initialised with a random walk, and TFs placed randomly within the simulation domain. Bead-bead overlaps are relaxed by first evolving the system for a few timesteps with a soft repulsive potential between beads. The system is then thermalized for $10^5 ~\tau_B$ with only repulsive Weeks-Chandler-Anderson potential interactions between all beads. Finally the system is evolved including attractive interactions and the observables of interest are monitored. A TU is considered transcriptionally active if the distance between its centre and the centre of almost one active TF is less than $2.23\sigma$. Averages are evaluated over $100$ independent runs for each case. Cluster are identified using the python-implemented DBSCAN algorithm \cite{dbscan} with distance threshold equal to the threshold for transcriptional activity.


\subsection{Mapping simulation units to physical units}

Simulation units can be readily converted into physical units. Using the realistic values of bead diameter $\sigma=3~10^{-8}~m$, room temperature $T=300~K$ and nucleoplasm viscosity $\eta_{sol}=10-100~cP$, we find $\epsilon\simeq4.14~10^{-21}~J$ and $\tau_{LJ}=\tau_B=\sigma^2/D=3\pi\eta_{sol}\sigma^3/\epsilon\simeq0.6-6~10^{-3}~s$.

\section{RESULTS}
\label{sec:results}

The present section is  devoted to the results of our investigation. We first report and comment the model phenomenology using a short chromatin filament,  and then we focus on human chromosomes and comparisons with experimental GRO-seq data.

\subsection{Short chromatin filament}
\label{sec:toy_model}

 As a first step, we consider a simple $M=1000$ beads long polymer chain. We investigate and compare many possible configurations varying the number of TFs in the system, the non-TU bead types (either non-binding or weakly-sticky) and the TU sequence. The TU bead sequence can be chosen in two different ways: either placing a fixed number of TUs in random positions, or placing consecutive TUs at fixed bead distance. Concerning the simulation parameter choice, the simulation side box is set to $L=100\sigma$, so that the the chain volume fraction is of order $\sim10^{-3}$, while the number $N$ of TFs is always set in such a way that their volume fraction is of order $\sim 10^{-6}$ (the exact number will be specified case by case). The switching rates are set at $\alpha_{off}=\alpha_{on}=10^{-5}~\tau_B$, so that in the steady state $N_a=N_p$ with $n_s=100\tau_B$, and the system is evolved for $8~10^5~\tau_B$.
 
\begin{figure}[ht!]
\centerline{\includegraphics[width=0.93\textwidth]{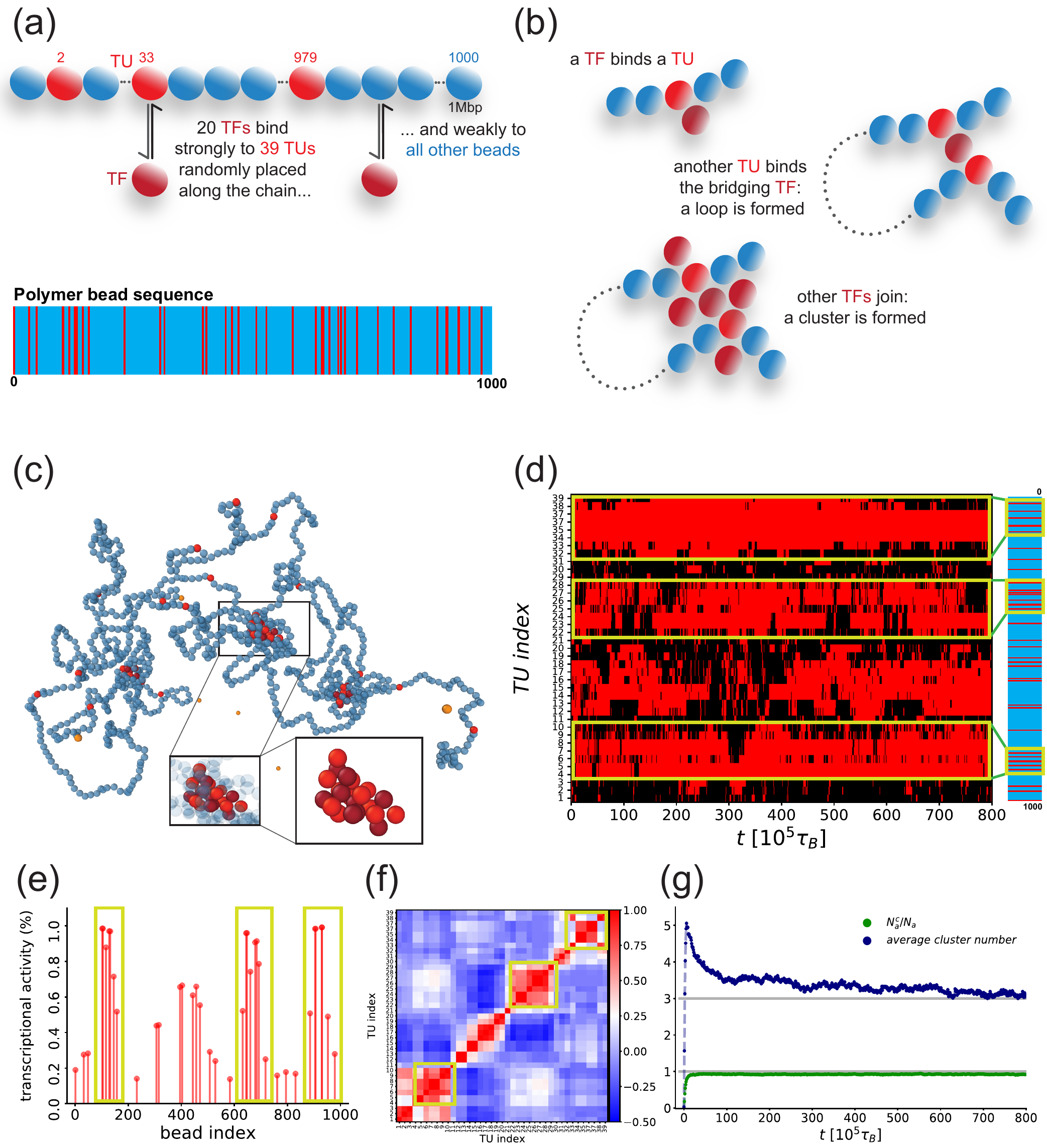}}
\caption{\footnotesize{\textbf{Reference case.} $\textbf{(a)}$: graphical depiction of the reference case and visual depiction of the polymer bead sequence. Active TFs (dark-red) switching between active/inactive states at rate $\alpha=10^{-5}~\tau_B^{-1}$ bind strongly to TUs (red) and weakly to all other beads (blue). \textbf{(b)}: bridging-induced attraction mechanism -- cluster formation stages. \textbf{(c)}: typical space conformation of the reference case bead polymer with active and inactive TFs (dark-red and orange, respectively) and enlargements of the boxed areas (inactive TFs not shown, weakly-binding beads depicted partially and completely transparent in the small and large inset, respectively). \textbf{(d)}: kymograph for a single significant simulation run lasting $8~10^{5}\tau_B$. The TU index denotes the sequential number each TU appears on the chain. Black and red pixels denote a transciptionally passive or active state, respectively. The green boxes highlight the transcriptional state of TUs involved in three stable clusters. The polymer bead sequence on the right offer visual aid in locating such TUs on the entire polymer bead sequence. $\textbf{(e)}$ and $\textbf{(f)}$: 
average transcriptional activity and Pearson correlation matrix. The green boxes highlight the cluster transcriptional activity and correlation values. $\textbf{(g)}$: cluster characterisation. The gray horizontal lines are a guide to the eye.
}}
\label{fig1}
\end{figure}

\textbf{Reference case.} The first case we considered is a chain with $39$ TUs placed randomly, the remainder polymer beads set as weakly-sticky and $N=40$ TFs ($N_a=N_p=20$). This case will be used in the following as reference for comparisons and therefore will be referred to as the {\it reference case}. A cartoon of the reference case is shown in \autoref{fig1}a, along with a visual depiction of the actual polymer bead sequence. The polymer model under scrutiny is characterised by a rich topological phenomenology\cite{brackley2016}: once a TU and a TF bind, it is more probable for other TUs to bind the same TF. The latter plays then the role of a bridge bewteen TUs and a chromatin loop forms. This makes in turn the local TU density increase, so that now it is more probable for other TFs to bind the same filament region, generating a cluster. This cycle repeats until the cluster reaches the maximum dimension allowed by the entropic cost for its formation. A visual depiction of the main stages of cluster formation is provided by \autoref{fig1}b. The phenomenon here described is a well known feedback mechanism called {\it bridging-induced attraction}. An actual typical fibre space conformation characterised by different clusters and loops is shown by \autoref{fig1}c. The insets report a close-up of the biggest cluster and underlines the involvement of a great number of TFs and TUs as well. 

In our model, the vicinity of a TU and almost one active TF at a given time is used to characterise that TU as transcriptionally active at that time. Significant observables to monitor are the transcriptional state over time of each TU (whether active or not), graphically depicted as a kymograph, and the average transcriptional activity of each TU, i.e. how long each TU has been active with respect to the total simulation time. \autoref{fig1}d shows the kymograph of a single typical run. Some of the TUs are essentially never transcribing, some others are always transcribing, and the remainder alternate between the two states. Interestingly, three continuous red stripes stand out from the black background. As highlighted by the polymer bead sequence and the green boxes, they are located in chain regions with higher local TU density, i.e. TUs are close to each other, which consistently sequester active TFs. As a consequence other TUs are prevented from binding to TFs, hence the transcriptionally passive and state-alternating TUs.

The cluster argument is supported by \autoref{fig1}e, reporting the average transcriptional activity for each TU as a function of the bead index. As highlighted by the green boxes, the activity profile shows in fact the appearance of three groups of high activity peaks, signalling the stable formation of the three aforementioned clusters in every run, alternated by low activity areas, corresponding to low transcribing TUs. Further support is provided by \autoref{fig1}f, reporting instead the transcriptional Pearson correlation matrix in which each $ij~,i,j=1,\ldots,n_{TU}$ pixel is coloured according to the Pearson correlation between the $i$-th and the $j$-th TUs. Cluster TUs participate to the same action (being transcribed) for long time periods so, as underlined by the green squares, they result strongly positively correlated. Oppositely, as cluster TUs are involved in the sequestering dynamics, couples of TUs one inside and one outside clusters perform for long time different actions (being and not being transcribed), so they result negatively correlated.

The time series of the average number of clusters and of the average fraction of active TFs in clusters with
respect to the total active ones $N^c_a/N_a$ reported in \autoref{fig1}g shows that the system reaches a stationary state in a relatively short time. At long times, the average cluster size slightly oscillates around $3$. This value is in line with our previous identification of three highly active TUs regions involved in as many clusters. The small fluctuations are instead due to the formation and breaking of small temporary clusters. The average fraction of active TFs in clusters reaches instead a value $\sim 1$. Combining the information that essentially all active TFs are occupied in transcriptional activities with the identification of three main clusters argued above, it appears clearer that TFs are actually segregated in such clusters and TUs undergo a non negligible competition to be expressed.

\begin{figure}[ht!]
\centerline{\includegraphics[width=1.0\textwidth]{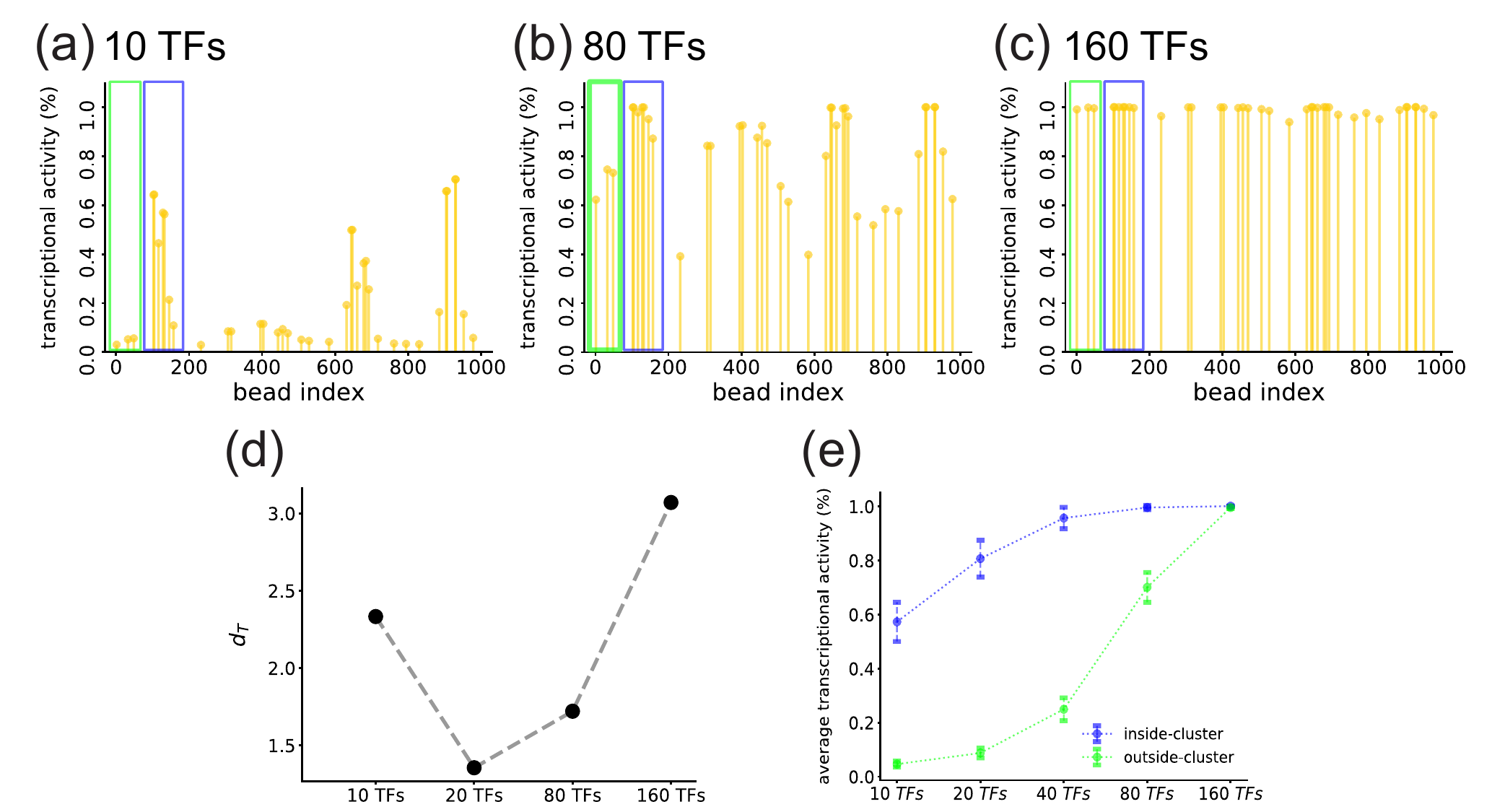}}
\caption{\footnotesize{\textbf{Variation of TFs number.} $\textbf{(a)}$,  $\textbf{(b)}$ and  $\textbf{(c)}$: average transcriptional activity for the reference case with $N=10$, $N=80$ and $N=160$ TFs, respectively. The green and blue boxes highlight the inside- and outside-cluster TUs we focus on in panel $\textbf{(e)}$. $\textbf{(d)}$: transcriptional difference between the random case with $N=40$ TFs and the cases with $N=10$, $N=20$, $N=80$ and $N=160$ TFs. $\textbf{(e)}$: mean average transcriptional activity for inside- and outside-cluster TUs as a function of the TFs number. Bars report the standards deviation. Points are coloured according to the boxes in panels $\textbf{(a)}$,  $\textbf{(b)}$ and $\textbf{(c)}$.}}
\label{fig2}
\end{figure}

\textbf{Variation of TFs number.} As a first model perturbation, we investigated the effects produced by the variation of the TFs number with respect to the reference case with $N=40$ TFs. \autoref{fig2}a, b and c report the transcriptional activity profile for the reference case with $N=10$ ($N_a=N_p=5$), $N=80$ ($N_a=N_p=40$) and $N=160$ ($N_a=N_p=80$) TFs, respectively. These panels immediately make it clear that as the number of TFs is increased, the activity profile shows the appearance of peaks in the same locations but with increasing value, up to saturation for $N=160$ TFs. 

In order to quantify how much difference there is between the transcriptional profiles from the reference case with $N=40$ TFs, and the ones with other values of $N$, a useful observable is the transcriptional difference $d_T$ \cite{brackley2021}. This quantity is defined as the euclidean distance between the activity vector $\{a'_i\},~i=1,\ldots,n_{TU}$ in one configuration and the activity vector $\{a_i\},~i=1,\ldots,n_{TU}$ in another one, in symbols
\begin{equation*}
    d_T=\sqrt{\sum_{i=1}^N(a'_i-a_i)^2}~.
\end{equation*}
 Simple algebraic arguments show that the transcriptional difference is minimum with value $d_T=0$ when the activity of all TUs is the same, while it is maximum with value $d_T=\sqrt{n_{TU}}$ when the activity of TUs in one case are all $0$ and the ones in the other case are all $1$. As intuitively expected, \autoref{fig2}d shows the highest $d_T$ values to occur for the two cases with maximum and minimum number of TFs considered, while the smallest one for the case with the closest number of TFs to the reference case. The TFs number variation therefore appears as a viable pathway to alter the transcriptional behaviour of a chromatin fibre, with the most remarkable changes obtained when the number of TUs is either so low as to make the mean transcriptional activity of each TU low or negligible (see \autoref{fig2}a), or so high as to saturate it (see \autoref{fig2}c). 
 
 More insight into the TF-mediated transcriptional regulation is provided by \autoref{fig2}e, which reports the mean transcriptional activity for the leftmost TUs respectively inside and outside clusters (blue and green boxes in \autoref{fig2}a, b and c) as a function of the TFs number. The figure shows that both groups of TUs increase their mean transcriptional values. Due to cluster formation and stability, inside-cluster TUs start from a value $\sim 0.6$ and reach the saturation value $\sim 1.0$ already at $N=40$ TFs with a concave trend. More repressed outside-cluster TUs start instead from a negligible value and slowly increase their mean activity until reaching saturation only for $N=160$ TFs thanks to the large availability of TFs.
 
\textbf{Role of weakly-binding beads.} 
\begin{figure}[t]
\centerline{\includegraphics[width=1.0\textwidth]{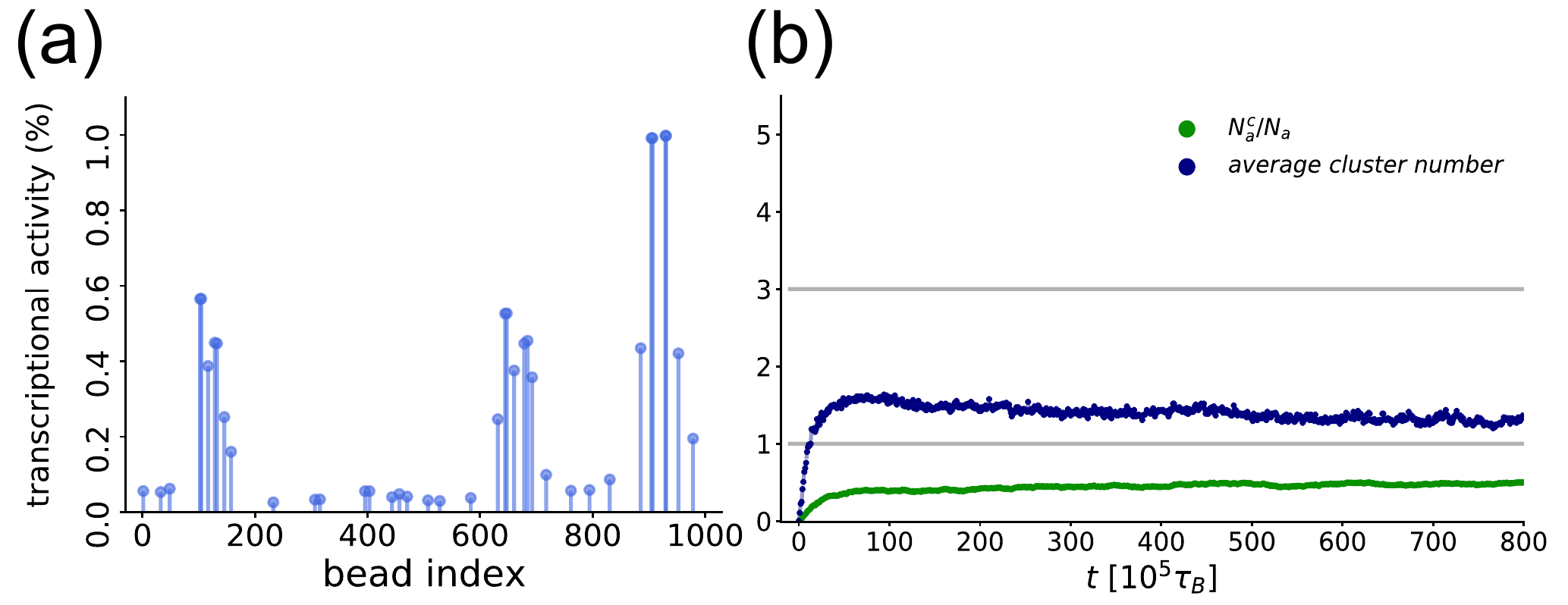}}
\caption{\footnotesize{\textbf{Role of weakly-binding beads.} $\textbf{(a)}$: average transcriptional activity for the reference case with non-binding beads instead of weakly-binding ones. $\textbf{(b)}$: cluster characterisation. The eye-aiding gray horizontal lines are the same as in \autoref{fig1}f.}}
\label{fig3}
\end{figure}
We then investigated the role played by the weakly-binding beads. This is done by substituting the weakly-binding beads in the reference polymer chain with non-binding ones, leaving the TF switching rates, the TUs position and the TUs and TFs number unchanged. \autoref{fig3}a shows this to have an important impact on the transcriptional profile: although the highest peaks appear in the same locations as in the reference case, they are much lower, and all other previous bars are now either absent or negligible. In terms of transcriptional difference, the overall configuration is similar to \autoref{fig2}a. For the non-binding beads case we in fact have $d_T\sim 1.76~10^3$ and for the reference case with $N=20$ TFs, where the dramatic transcription reduction was due to a $75\%$ decrease in the TFs number, $d_T\sim 1.36~10^3$. It is then natural to conjecture that the weakly-binding beads play an important role, along with the TFs, in cluster formation and stability. A first support to this hypothesis comes from the visual inspection of a number of non-binding case configurations. We in fact observe the formation of smaller clusters, breaking and reforming during the system evolution. A more quantitative and conclusive support is provided by the average fraction of active TFs in clusters with respect to the total active ones $N_a^C/N_a$ and average number of clusters time series reported in \autoref{fig3}b. In contrast to the value $\sim 1$ from the reference case, the average fraction of TFs in clusters reaches the constant value $\sim 0.5$. In other words, on average only half of the active TFs are now involved in clusters, with a consequent much smaller average cluster dimension. The average cluster number is also strongly affected as it decreases from $\sim 3$ in the reference case to $\sim 1.3$. This value is larger than $1$ beacuse it takes into account the cluster involving the TU with mean trancriptional activity $\sim 1$ around bead index $900$ plus some smaller temporary clusters quickly forming and breaking. From these data we infer that: (i) the dramatic role of weakly-binding beads lies in their ability to glue together TFs and TUs in clusters, allowing them to be larger in number and dimension and not break when some state-changing TF leaves the cluster; (ii) both TF number and chain bead type alterations result in similar and significant transcriptional difference values.

\textbf{Regular TU pattern.} \begin{figure}[t]
\centerline{\includegraphics[width=1.0\textwidth]{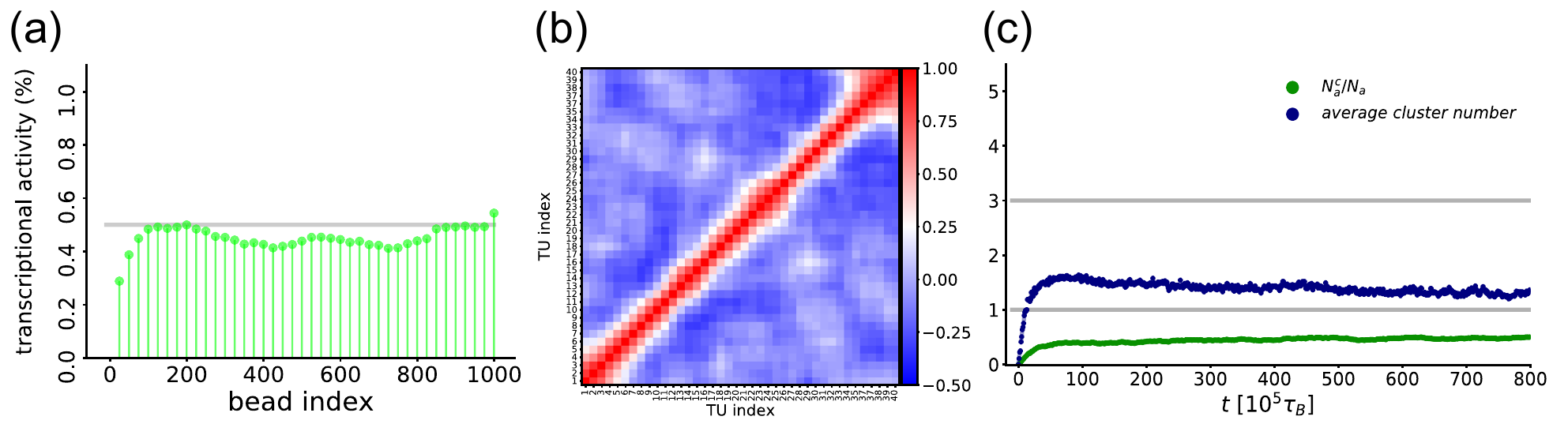}}
\caption{\footnotesize{\textbf{Regular TU pattern.} $\textbf{(a)}$: average transcriptional activity for the regular TU pattern case. $\textbf{(b)}$: Pearson correlation matrix. $\textbf{(d)}$: cluster characterisation. The gray horizontal lines are a guide to the eye.}}
\label{fig4}
\end{figure}
We also investigated the role of the sequence of TU beads in cluster formation and activity profile by simulating a chain whose consecutive TU beads are regularly spaced every $n_{sp}$ beads. The polymer length and TFs number are kept fixed at $M=1000$ beads and $N=40$ ($N_a=N_p=20$) respectively. In order to simulate a case with a number of TUs close to the reference case, the TU bead distance is set at $n_{sp}=25$, resulting in $n_{TU}=40$ TUs.

\autoref{fig4}a reports the average transcriptional activity for this case, which as expected now appears flat, around the value $\sim 0.5$. TUs at the two ends of the filaments experience boundary effects: the left-most ones are shielded by weakly-binding beads, so it is a bit more difficult for them to access TFs, resulting in a slightly lower activity; the rightmost one is instead the very last bead of the filament, so TFs can bind it a bit more easily, hence its slightly higher activity. Interestingly, the flat transcriptional value can be modulated acting on the system parameters. Thus, setting $\alpha_{off}=10^{-5}~\tau_B^{-1}$ and $\alpha_{on}/\alpha_{off}=0.25$ so that $N_a=8, N_p=32$ leads to $\sim 0.25$, while keeping fixed the switching rates and setting $N=80$ TFs leads to $\sim 0.75$.

The shape of the activity profile is mirrored by the TU correlation. As shown by the Pearson correlation matrix in \autoref{fig4}b, consecutive or close TUs are positively correlated since they tend to be in the same cluster for a large fraction of time. Far apart TUs instead do not influence each other as much as in \autoref{fig1}f and result in very low reciprocal correlation values.

As the average cluster number time series in \autoref{fig4}c signals the formation of clusters, the only possibility to reconcile this observation with the flat transcriptional profile is that clusters can form with the same probability in every position along the polymer chain. This phenomenon is clearly driven by the regular TU bead sequence and supported by visual inspection of the activity profiles of single runs. In the reference case, where the only difference is the TU bead sequence, this phenomenon in fact does not occur. Here there are in fact groups of close consecutive TUs distinctly separated from other close groups of close TUs, each of them facilitating the formation of the same separate clusters in every run. We can therefore assert that the variation in the TU bead sequence is yet another pathway to regulate the topological features of the system regarding the formation of the same clusters in each run: as the TU spacing changes from uniform to random with groups of close TUs, the appearance of clusters becomes essentially deterministic, and consequently the transcripitonal profile transitions from flat to inhomogeneous, and characterised by groups of peaks. 

\begin{figure}[ht!]
\centerline{\includegraphics[width=1.0\textwidth]{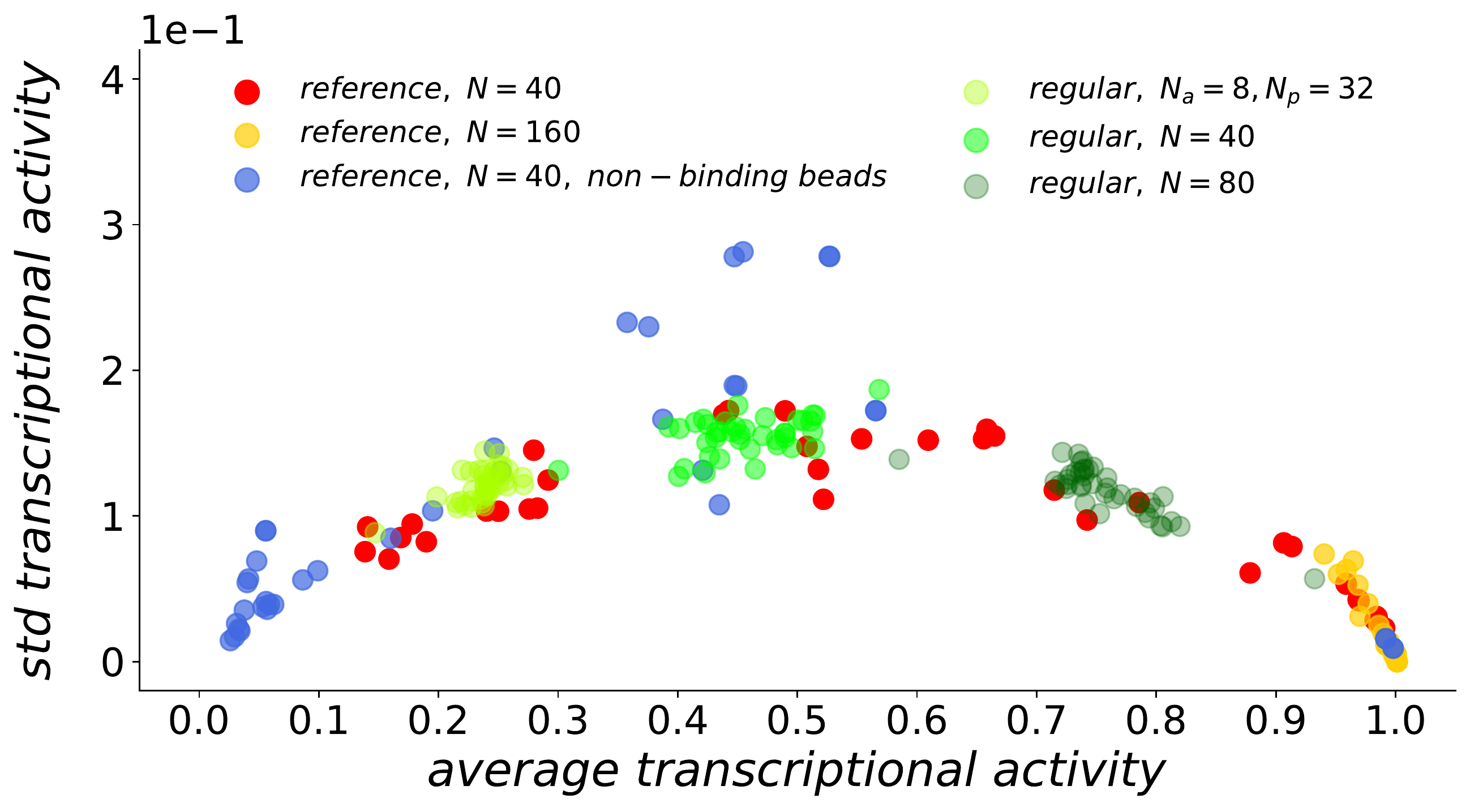}}
\caption{\footnotesize{\textbf{Statistics of transcriptional activity.} Plot of the standard deviation of the transcriptional activity as a function of the average of the same observable for some significant cases. Each point represents the couple transcription average-standard deviation for a single TU.}}
\label{fig5}
\end{figure}

\textbf{Statistics of transcriptional activity.} Finally, we make some comments about the transcriptional activity statistics. \autoref{fig5} reports the transcriptional activity standard deviation (std) as a function of the average of the same observable for each TU for some significant cases. Data points roughly follow a parabolic trend. The reference case points span the largest std and average intervals and best depict the parabolic shape. Note that the maximum std is reached for average values $\sim 0.5$, meaning that the TU resulting in such an average are those for which the std, hence transcriptional fluctuations, are the largest. Moving towards average transcriptional activity $0$ and $1$, the std reduces symmetrically. Together with the transcriptional profile results, this suggests that the TU bead sequence along the polymer chain induces topological conformations such that some TUs are saturated or non transcribed with a certain stability across different runs, thus having almost zero std. 

The occurrence of high stds for transcriptionally fluctuating TUs is confirmed by looking at the regular TU pattern cases. As reported above, and as is clear from \autoref{fig5}, in these cases TUs on average are regularly transcribed with essentially the same parameter-dependent average transcriptional activity values. Their corresponding stds are comparatively high and consistent with the reference case TUs ones with similar average values. The characterisation for the extrema mean values $1$ and $0$ is instead better explained by looking at the case with $N=160$ TFs and the case with non-binding beads instead of weakly-sticky. The former is in fact characterised by a transcriptional profile with almost all TUs saturated (see \autoref{fig2}c). As such TUs essentially perform very similarly in each run, their stds are clearly low, and as shown by the figure, much lower than data points with average $\sim 0.5$. The hypothetical limiting case of a deterministic TU with transcriptional activity $1$ in each run would lead to an std of $0$, and thus complete certainty about its transcriptional outcome. In the latter case we have instead the opposite situation of most TUs non transcribed, and thus with average $\sim 0$ and low fluctuations. As shown by \autoref{fig3}a, the remainder TUs are characterised by considerable average transcriptional activities and stds, which seems to hint to an approximately parabolic trend. Further simulations with non-binding beads and different number of TFs support the trend emerging from the reference case. 

\subsection{Human chromosomes}
\label{sec:chrs}

\begin{figure}[ht!]
\centerline{\includegraphics[width=1.0\textwidth]{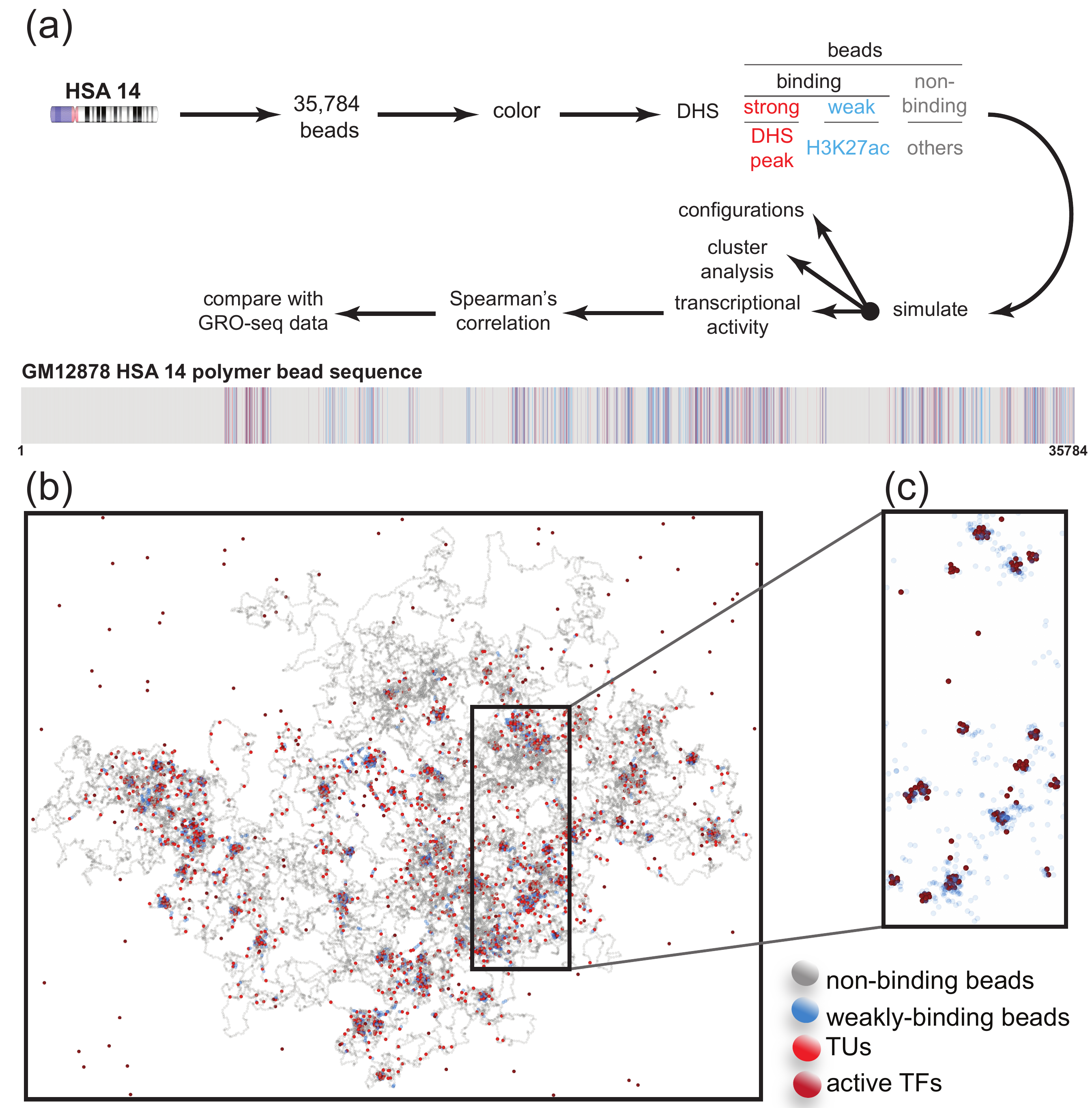}}
\caption{\footnotesize{\textbf{GM12878 HSA $14$ configuration.} $\textbf{(a):}$ DHS model workflow and scheme of the GM12878 HSA $14$ polymer bead sequence. Bars denote the bead position and are coloured according to the legend at the bottom right corner.  $\textbf{(b):}$ typical space conformation of the GM12878 HSA $14$ polymer chain. Beads are coloured according to their types as reported by the legend (inactive TFs not shown). $\textbf{(c):}$ close-up of a portion of the entire space configuration (inactive TFs and non-binding beads not shown).}}
\label{fig6}
\end{figure}

We now turn to the simulation of human chromosomes. In this section our goal is to  validate the predictive power and reliability of the diffusing trascription factors model by comparing its numerical outputs with GRO-seq experimental data across different chromosomes and cell lines. In particular, we study HSA $14$, HSA $18$ and HSA $19$ in human umbilical vein endothelial cell (HUVEC) line; we also study HSA $14$ in B-lmphocyte GM12878 cell line. Our choice falls on HSA $14$ because it is an intermediate size human chromosome and it 
has been the object of a number of recent studies hence is well characterised by simulations which can be used as a benchmark~\cite{brackley2013, brackley2021}. The choice for the other two HUVEC chromosomes is instead driven by their gene densities. HSA $18$ is in fact an example of gene-poor chromosome as, in polymer terms, only $\sim 9\%$ of the beads are either weakly-sticky or TUs ($\sim 3\%$ weakly-sticky, $\sim 6\%$ TUs, see below for workflow to determine weakly-sticky beads and TUs). HSA $19$ is instead an example of gene-rich chromosome, as $\sim 17\%$ of beads are binding ($\sim 6\%$ weakly-sticky, $\sim 11\%$ TUs). HSA $14$ is instead in an intermediate ranking, as $\sim 10\%$ of its beads are binding ($\sim 3\%$ weakly-sticky, $\sim 7\%$ TUs). 

For each chromosome, the bead polymer sequence is built following the DHS model workflow\cite{brackley2021} schematised in \autoref{fig6}a for the GM12878 HSA$14$. The starting point are the ChIP-seq for H3K27ac and DNase-hypersensitivity (DHS) data provided by ENCODE \cite{encode}. The former data are optimal for showing chromosome sites that significantly correlate with open chromatin, i.e. to locate weakly-binding beads, while the latter data can be used to locate promoters and enhancers, i.e. the TUs position. Experimental data are then coarse-grained into the chosen base-pair resolution. The type of each chromosome bead is determined in the following way: beads containing a peak in H3K27ac but not in DHS are marked as weakly-sticky, beads containing a peak in DHS are marked as TUs, and all other beads are marked as non-binding. The extracted bead polymer sequence schematised in \autoref{fig6}a is used to start the simulations. The box side length $L$, TFs number $N$ and polymer bead length $M$ will be specified case by case. However we anticipate that $L$ is always chosen so that in all cases the chain volume fraction is of order $\sim10^{-3}$ and the TFs volume fraction is of order $\sim 10^{-4}$. Note that both chromatin and TF have higher density with respect to the short chromatin filament case from \autoref{sec:toy_model}: although the ratio between active TF and TU numbers remains similar, the higher concentration, as detailed below, leads to some difference in the qualitative behaviour.  
The switching rates are set at $\alpha_{off}=10^{-5}~\tau_B$ and $\alpha_{on}=\alpha_{off}/4$ with $n_s=100\tau_B$, so that in the steady state $N_a/N_p=0.25$. This choice is inspired by the literature \cite{brackley2021} and reproduces the experimentally observed $N_a/N_p$ ratio \cite{cook2001}. The system is evolved for $2~10^5~\tau_B$.

\subsubsection{HSA $14$ for GM12878 and HUVEC cell lines}
\label{sec:same_hsa}

To start, we focused on simulations of the same chromosome, HSA $14$, from the two cell lines GM12878 and HUVEC. The polymer is $M=35784$ beads long with $n_{TU}=2226$ TUs, the simulation side box side is set to $L=570\sigma$ and the TFs number to $N=5376$ TFs ($N_a=1075,N_p=4301$). \autoref{fig6}b reports a typical space conformation of the GM12878 chromosome (similar comments apply to the HUVEC one) formed by all possible types of polymer beads: non-binding beads, weakly-binding beads and TUs. The configuration reproduces features similar to the ones observed for the short chromatin filament of \autoref{sec:toy_model}, in particular we observe the formation of chromatin loops and clusters of various dimensions driven by the bridging-induced attraction feedback mechanism.
These are clearly visible from the configuration close-up reported in \autoref{fig6}c, which at the same time better highlights the gluing role played by weakly-binding beads inside clusters discussed above.


\begin{figure}[ht!]
\centerline{\includegraphics[width=1.0\textwidth]{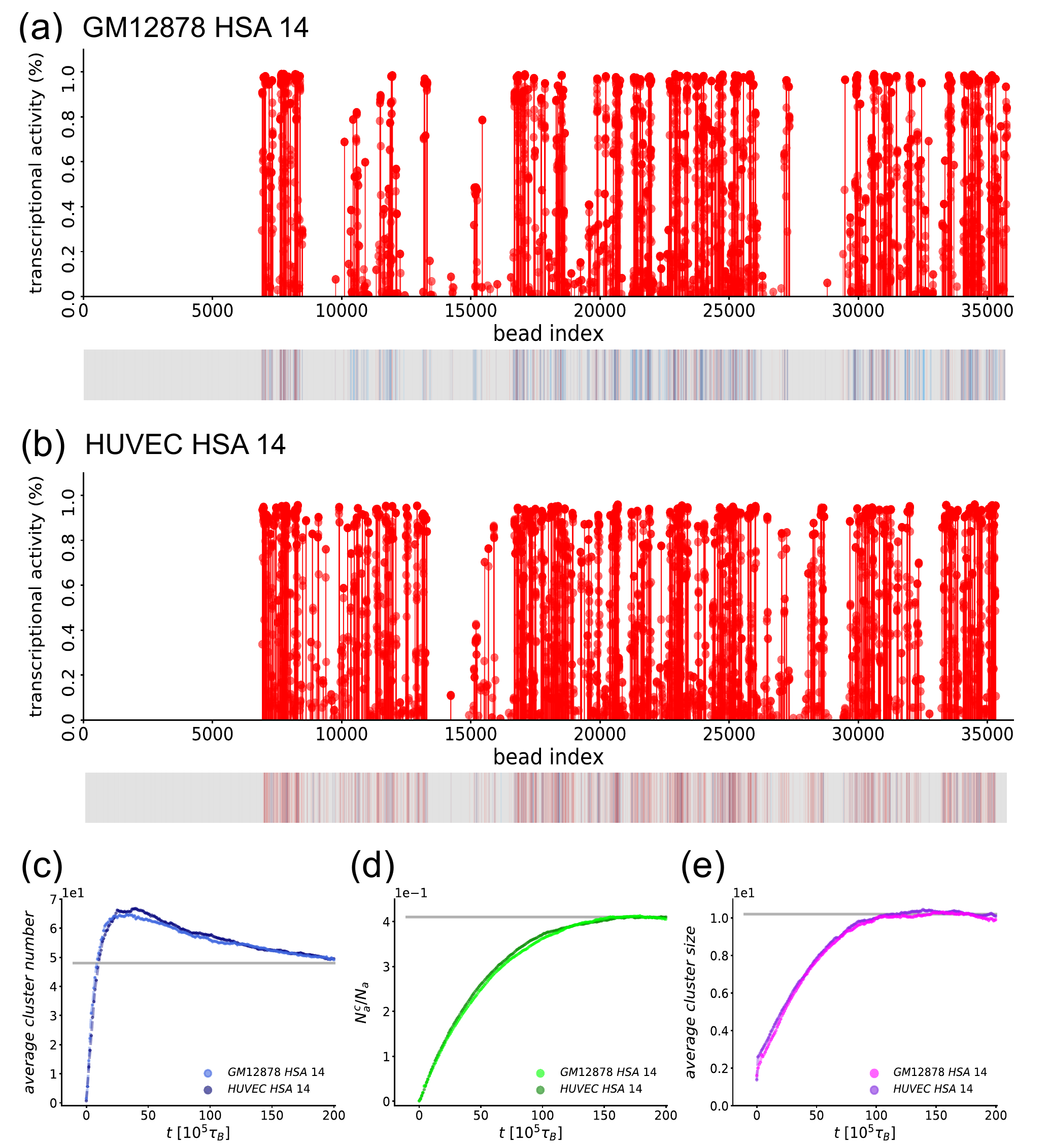}}
\caption{\footnotesize{\textbf{GM12878 and HUVEC HSA $14$.} $\textbf{(a)}$ and $\textbf{(b)}$: average transcriptional activity for GM12878 and HUVEC HSA $14$, respectively. The bottom part of the panels reports the chromosome polymer bead sequence generated by the DHS model. $\textbf{(c)}$, $\textbf{(d)}$ and $\textbf{(e):}$ time series for average cluster number, $N_a^C/N_a$ and average cluster size, for both cell types, respectively.}}
\label{fig7}
\end{figure}


\autoref{fig7}a and b report the average transcriptional activity for GM12878 and HUVEC HSA $14$, respectively, along with the polymer bead sequences produced by the DHS model. Both cases show regions with zero transcriptional activity, indicating the total absence of TUs in such areas, and regions with high activity peaks, indicating the formation of clusters in such other areas. The overall transcriptional profile is similar in both cases, with slightly lower transcriptional activity values for HUVEC in regions with low local TUs density. Concerning the cluster analysis, \autoref{fig7}c, d and e show the time series of the average cluster number, of the average fraction of active TFs in clusters with respect to the total active ones $N_a^c/N_a$ and of the average cluster size. These plots have the same trends and reach essentially the same stationary values in both cell lines. This is a first sign of the model reliability over different cell lines. The stationary average cluster number is $\sim 50$, while $N_a^c/N_a\sim 0.4$, corresponding to $\sim 430$ active TFs in clusters. The information that, as also shown by \autoref{fig7}a, cluster TUs are almost transcriptionally saturated and that a large fraction of active TFs is not bound at all to them suggests that the chosen $N_a$ is such that essentially all possible potential clusters have actually formed. Consequently in this case the sequestering dynamics discussed in \autoref{sec:toy_model} plays a minor role and the transcriptional competition between clusters results is diminished. Interestingly, the stationary average cluster size $\sim 10$ shows a considerable amount of active TFs to be involved in each cluster.

\begin{figure}[ht!]
\centerline{\includegraphics[width=0.75\textwidth]{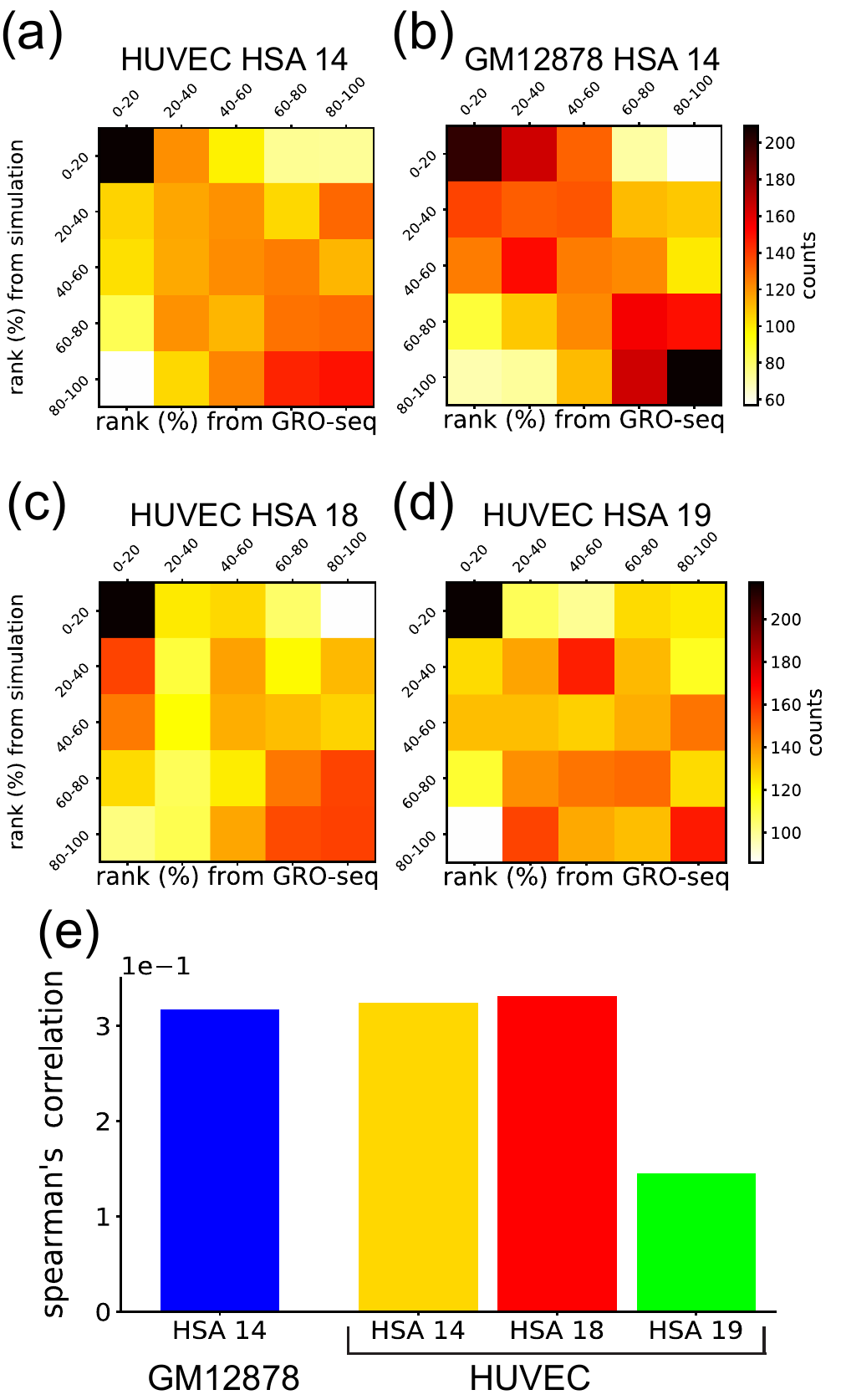}}
\caption{\footnotesize{\textbf{Chromosomes transcriptional overview.} $\textbf{(a)}$, $\textbf{(b)}$, $\textbf{(c)}$ and $\textbf{(d)}:$ comparison between numerical transcriptional activity data and experimental GRO-seq data ranked from $0-100\%$ for HUVEC HSA $14$, GM12878 HSA $14$ and HUVEC HSA $18$ and $19$, respectively. $\textbf{(e):}$ comparison between transcriptional activities from simulations and GRO-seq data trough Spearman’s rank correlation coefficient evaluation.}}
\label{fig9}
\end{figure}

In order to monitor the model performance, we compare our numerical results with  experimental GRO-seq data. A first qualitative indication pointing towards the validity and reliability of the model is provided by \autoref{fig9}a and b, reporting a comparison between numerical transcriptional activity data and experimental GRO-seq data ranked from $0-100\%$ for HUVEC and GM12878 HSA $14$, respectively. As a direct comparison is not feasible the idea here is to extract from both the numerical and the experimental transcriptional data distributions the quintile values, then rank each TU numerical and experimental transcriptional activity with respect to such quintile intervals (first quintile $0-20\%$, second quintile $20-40\%$, and so on up to the fifth quintile $80-100\%$), and finally counting the occurrences of all possible couple of rankings across all TUs. Strong values over the plot diagonals would denote a significant correspondence between ranked data, and this is exactly what the two panels show. A more quantitative proof is provided by the Spearman's rank correlation coefficient\cite{jerome2011} between the same sets of numerical and experimental data, reported in \autoref{fig9}e and found to be on average $\sim 0.32$ with an estimated uncertainty of order $10^{-3}$ in both cases (mean and standard deviation obtained using bootstrap with $100$ resamplings). The correlation values themselves denote a significant correlation between numerical and experimental data, while the occurrence of essentially the same values amongst different cell lines, is yet another sign of the fact that the model performs similarly in different biological conditions. 

\subsubsection{HSA $18$ and $19$ from HUVEC cell line}
\label{sec:same_cell}

To further test the model, we then fixed the cell line and considered different chromosomes. As aforementioned, we focused on HSA $18$ and $19$ from HUVEC. 

\begin{figure}[ht!]
\centerline{\includegraphics[width=1.0\textwidth]{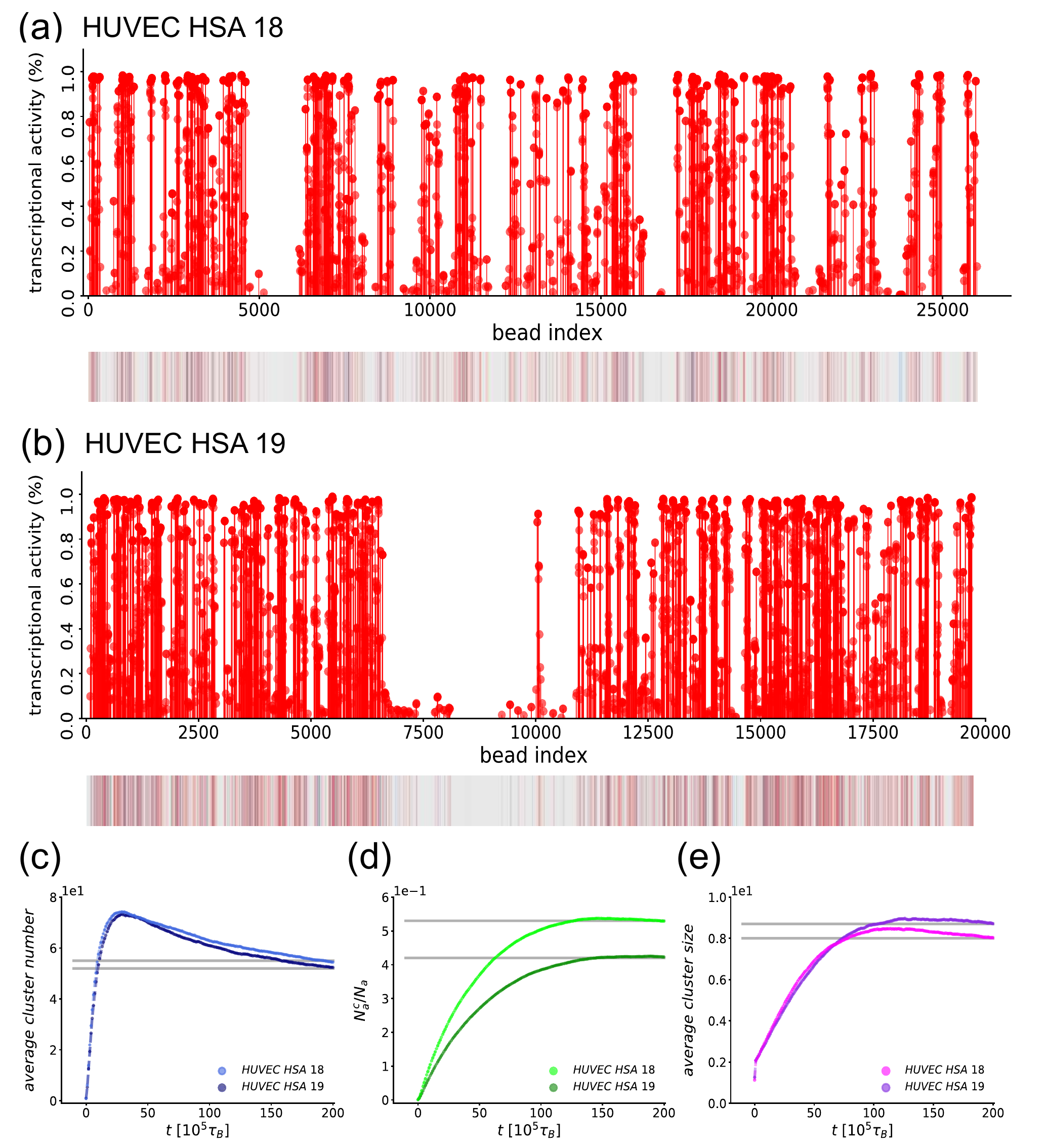}}
\caption{\footnotesize{\textbf{HUVEC HSA $18$ and HSA $19$.} $\textbf{(a)}$ and $\textbf{(b)}$: average transcriptional activity for HUVEC HSA $18$ and HSA $19$. The bottom part of the panels reports the chromosome polymer bead generated by the DHS model. $\textbf{(c)}$, $\textbf{(d)}$ and $\textbf{(e)}$: time series for average cluster number, $N_a^C/N_a$ and average cluster size, for both chromosomes, respectively.}}
\label{fig8}
\end{figure}

For HSA $18$, the polymer is $M=26026$ beads long with $n_{TU}=2226$ TUs, the simulation side box side is set to $L=500\sigma$ and the TFs number to $N=1832$ TFs ($N_a=733, N_p=2931$). For HSA $19$, the polymer is instead $M=19710$ beads long with $n_{TU}=2117$ TUs, the simulation side box is set to $L=556\sigma$ and the TFs number to $N=5052$ ($N_a=1010, N_p=4042$). \autoref{fig8}a and b report the average transcriptional activity for HSA $18$ and $19$, respectively. As for HSA $14$, both transcriptional profiles follow the TU disposition along the chain, with TUs in cluster almost saturated. 

Concerning the cluster features, \autoref{fig8} c, d and e show that also for HUVEC HSA $18$ and $19$ the average cluster number, $N_a^c/N_a$ and average cluster size time series have similar trends, in turn similar to the HSA $14$ one from \autoref{fig7}. The novelty here is that while the stationary values of cluster number and average cluster size are of the same order of magnitude, there tend to be slightly more clusters but smaller in size with respect to the case of HSA $14$. 

\begin{figure}[ht!]
\centerline{\includegraphics[width=1.0\textwidth]{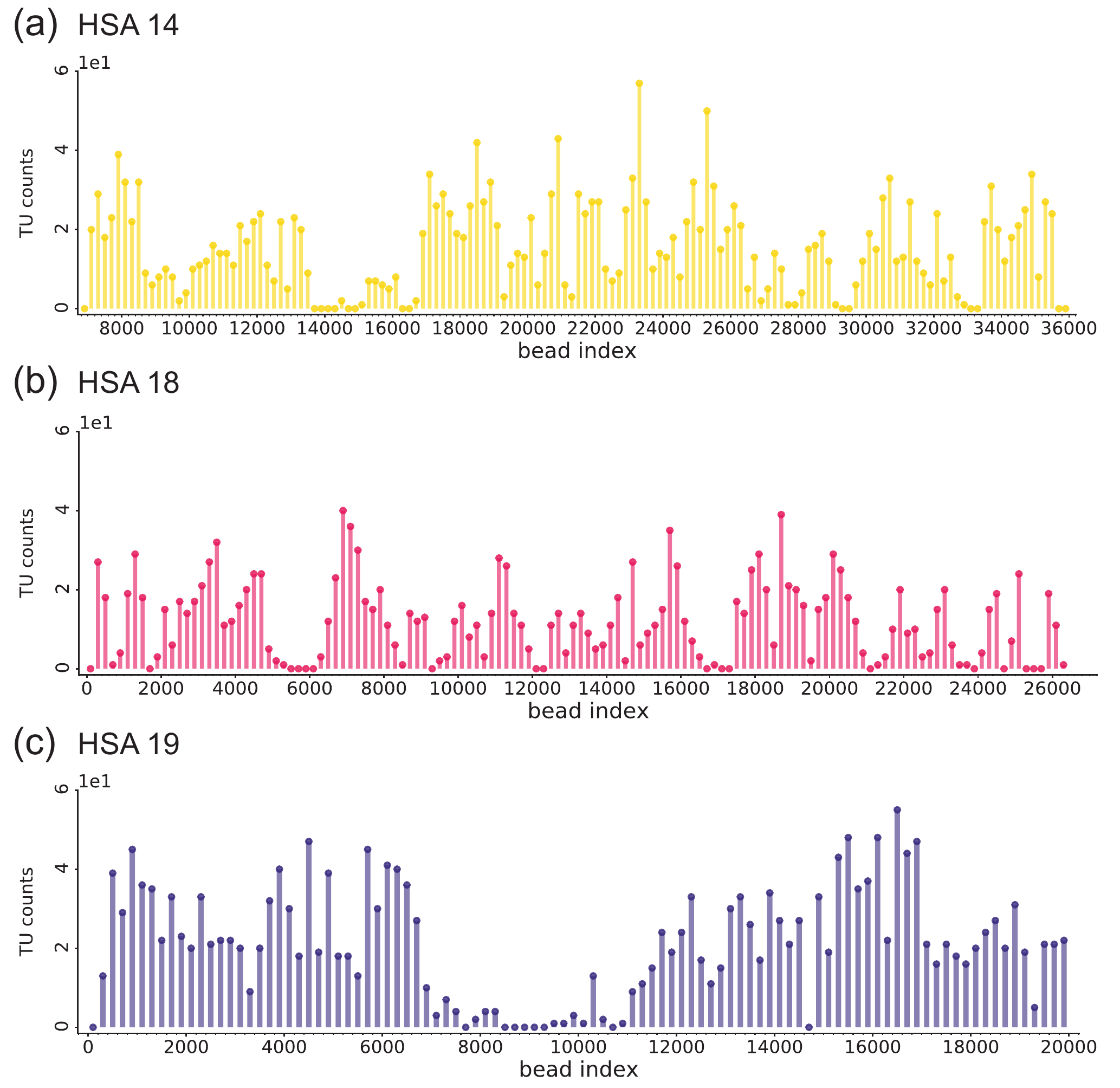}}
\caption{\footnotesize{\textbf{Chromosome TU counts profiles}. $\textbf{(a)}$, $\textbf{(b)}$ and  $\textbf{(c)}$: TU counts profile along the polymer chain for HSA $14$, HSA $18$ and HSA $19$, respectively. In $\textbf{(a)}$ we cut the initial $6800$ beads as no TUs appear there (see \autoref{fig7}a). Small ticks denote the $200$ beads bins width.}}
\label{fig10}
\end{figure}

Concerning the model performance, for HSA $18$ the situation is essentially similar to the HSA $14$ cases, even if some high off-diagonal values appear. \autoref{fig9}c shows in fact an overall good qualitative agreement between numerical and experimental data. However, the model results are validated by the Spearman's rank correlation coefficient, found to be $\sim 0.36$ on average with an estimated uncertainty of order $10^{-3}$ and graphically reported in \autoref{fig9}e. Such a measurement is completely consistent with the one concerning HSA $14$, and is then a proof of the model reliability also across chromosomes. For HSA $19$ the situation is instead a bit worse. We in fact notice from \autoref{fig9}d that the ranking comparison count plot is flatter with many of the off-diagonal counts similar to the diagonal ones. Moreover, the Spearman's rank correlation coefficient is now on average $\sim 0.15$ with an estimated uncertainty of order $10^{-3}$. This result is yet statistically significant, but clearly much lower than in previous cases. A possible reason explaining this finding could be related to the DHS model mechanism. We recall that promoters directly bind proteins to initiate transcription, while enhancers can bind proteins to increase the probability that a specific gene is transcribed. In our machinery, DHS data, marking promoters and enhancers as well, are used by the DHS model to identify TU beads, whose sticky action should mimic the biological functioning of both promoters and enhancers. The discrepancy between numerical results and experimental data could therefore be traced back to the fact that our model essentially neglects the biochemical processes involved in the enhancers behaviour, treating the latter {\it de facto} as promoters. A logical consequence is that the model performance should worsen as i) the global chromosome gene density increases; ii) TUs become more evenly distributed. Point i) takes into consideration the mere fact that the more the TUs are, the more the ill-modelled enhancers. Point ii) takes instead into account that a close group of TUs, coarse graining promoters or enhancers or both, leads anyway to the formation of a cluster, while as seen in \autoref{fig4}, many almost evenly spaced TUs in the model would lead to the regular formation of clusters in every position of such region,
neglecting the bias provided by the exact location of promoters and enhancers. For the sake of clarity, \autoref{fig10} reports the TU counts profiles for the chromosomes of interest obtained by binning an interval of the same bead length of each polymer chromosome and counting the TU occurrences in each bin. Such profiles follow closely the average transcriptional profiles from \autoref{fig7}a and b and \autoref{fig8}a and b and show that different global TU densities are in this case associated to counts (transcriptional) profiles with different features: in the HSA $19$ case, apart from the central region, the counts profile shows many consecutive high and similar counts values, while in the HSA $18$ case the profile shows high counts peaks alternated by many low-counts regions. Coming back to the results, the hypothesised gene density and disposition correlation worsening seems to be confirmed by our data: the model in fact performs worst in the gene-rich and similar-counts HSA $19$ case and best in the gene-poor peaked-counts HSA $18$ case. 

\section{CONCLUSIONS}
\label{sec:conclusions}

In this paper we have performed molecular dynamics simulations adopting the diffusing transcription factors model to study chromatin 3D conformations and transcriptional processes. The model is first employed on a short chromatin filament to highlight some generic emerging properties of the model. 
We then performed simulations of entire human chromosomes to prove the validity of the model predictions for transcriptional activity. 

Concerning the toy short filament model, we considered various systems versions fixing the polymer bead length and varying either the number of TFs, or the polymer TUs, weakly-sticky and non-binding bead sequence. To start, we considered a reference case where the TUs are randomly positioned and the remaining polymer beads are weakly-sticky. We reproduced some known results, such as the cluster formation driven by the bridging-induced attraction feedback mechanism and a non-uniform transcriptional activity profile affected by the formation of such clusters. We then considered a number of perturbations and highlighted the main notable differences. First, we varied the TFs number in the system whilst keeping the polymer chain unchanged, and proved the transcription factor concentration is important for gene expression, as one can transition from a regime with negligible transcription to a  saturation regime where all TUs are transcribed most of the time, as the TF count is increased. Second, we explored the role of the weakly-binding beads by mutating all such kinds of beads in the reference case into non-binding ones. In this way we were able to show that weakly-sticky beads play a major role in cluster formation, size and stability. 
Third, we showed the TUS sequence, i.e. local TU density, to have a significant impact on the overall model output. By simulating a short polymer chain with the same bead length as the reference case one and consecutive TUs regularly spaced, we obtained a flat average transcriptional profile. We traced back such a feature to the formation of clusters in random positions along the chain, with complete loss of TU-specificity in the predicted expression profile. Finally, we remarked a notable parabolic-shaped trend of the standard deviation transcriptional activity as a function of the average of the same observable. As a result of the chosen polymer bead sequence, switching rates and TFs number, some TUs have highly fluctuating transcription, while some others are forced into saturated or non-transcribing configurations characterised by very low standard deviation in transcription. Therefore, our results show that both chromatin topology and transcription can be regulated acting on both the transcription factor concentration and on the patterning of TUs and weakly-binding or non-binding beads.

Concerning the human chromosomes simulations, we considered HSA $14$, $18$ and $19$ in HUVEC cells, and HSA $14$ in GM12878 cells, so as to test the model performance on different chromosomes along the same cell-line and also along the same chromosome in different cell lines. We found that correlation between numerical and experimental data is statistically significant in all cases, in line with previous literature results. However, we observe the model to perform better as the chromosome gene density decreases. We hypothesise this to be a consequence of the chromosome chain building procedure, as it neglects some biochemical processes involved in enhancer and promoter functioning. 

Future work will need to address these issues related to gene-rich chromosome simulation, and in general should be focused on further improving the correlation between numerical and experimental data. 
A first step could be the inclusion in the model of the difference between enhancer and promoters. Other potential avenues to explore include instead the generalisation to the case of multiple types of transcription factors and units, so as to take into account the action of different proteins on different genes, and the inclusion of dynamic modification of transcription units, for instance to take into account known repression pathways \cite{ijms161226074}.

\section*{Acknowledgements}

This work was possible thank to the access to Bari ReCaS e-Infrastructure funded by MIUR through PON Research and Competitiveness 2007-2013 Call 254 Action I. We acknowledge funding from MIUR Project No. PRIN 2020/PFCXPE.

\bibliographystyle{IEEEtran} 
\bibliography{References.bib}

\end{document}